\documentclass{article}
\usepackage{amsmath}
\usepackage{graphicx}
\usepackage[normalem]{ulem}
\usepackage{moreverb}
\usepackage{setspace} 
\usepackage{authblk}
\usepackage{color}

\begin{document}

\singlespacing
\title{Resistive wall mode and neoclassical tearing mode coupling in rotating tokamak plasmas}
\author{Rachel McAdams$^{1,2}$}
\author{ H R Wilson$^{1}$}
\author{I T Chapman$^{2}$}
\affil{$^1$York Plasma Institute, Department of Physics, University of York, Heslington, York, YO10 5DD, UK}
\affil{$^2$EURATOM/CCFE Fusion Association, Culham Science Centre, Abingdon, Oxon, OX14 3DB, UK}
\date{}
\maketitle

\begin{abstract}
A model system of equations has been derived to describe a toroidally rotating tokamak plasma, unstable to Resistive Wall Modes (RWMs) and metastable to Neoclassical Tearing Modes (NTMs), using a linear RWM model and a nonlinear NTM model. If no wall is present, the NTM growth shows the typical threshold/saturation island widths, whereas a linearly unstable kink mode grows exponentially in this model plasma system. When a resistive wall is present, the growth of the linearly unstable RWM is accelerated by an unstable island: a form of coupled RWM-NTM mode. Crucially, this coupled system has no threshold island width, giving the impression of a triggerless NTM, observed in high beta tokamak discharges. Increasing plasma rotation at the island location can mitigate its growth, decoupling the modes to yield a conventional RWM with no threshold width.
\end{abstract}

\section{Introduction}
 To operate ITER \cite{ITER1999}  in Advanced Tokamak (AT) scenarios, which are desirable for high fusion gain, control of performance limiting magnetohydrodynamic (MHD) instabilities must be achieved.  AT scenarios are characterised by high bootstrap fraction and high $\beta_N=\beta$(\%)$/(I/aB)$, $\beta=2 \mu_0 \langle p \rangle / \langle B^2 \rangle$, (where $\langle p \rangle$ is volume averaged pressure, $B$(T) is external magnetic field, $I$(MA) is toroidal current, $a$(m) is minor radius)\cite{Kessel1994}. 
Neoclassical Tearing Modes (NTMs)\cite{Carrera1986} are one such performance limiting instability \cite{Hender2007}, with a detrimental effect on the achievable $\beta_N$. NTMs are unstable if two criteria are satisfied: $\beta$ must be sufficiently high, and the plasma must typically be "seeded" with a magnetic island produced by another activity \cite{Gerhardt2009,Haye2006}, such as sawteeth{\cite{Sauter2002, Chapman2010,Haye2000a, Gude1999a}}, fishbones \cite{Gude1999a} or ELMs \cite{Haye2000b}. This initial seed island will only grow, and subsequently saturate, if it exceeds a certain critical width \cite{Chang1995, Fitzpatrick1995, Wilson1996}. The flattening of the pressure profile caused by the presence of the island, and the consequential removal of the pressure gradient-driven bootstrap current in the vicinity of the magnetic island O-point enhances filamentation of the current density, providing a feedback mechanism for continued island growth. However, NTMs which do not appear to have grown from a seed island, so-called \emph{triggerless} NTMs, have also been observed experimentally \cite{Gude1999a, Kislov2001, Brennan2003, Garofalo1999}, usually associated with high $\beta$ {or near the RWM limit \cite{Maget2010}}.
Another instability which is responsible for limiting tokamak performance at $\beta_N>\beta_N^{no-wall}$ is the Resistive Wall Mode (RWM); a branch of the long-wavelength external kink mode, driven in a tokamak by pressure but which has a growth rate slowed by the presence of finite resistivity in surrounding conducting walls. RWMs are slowly growing on the timescale of the wall diffusion time, and have a complex relationship with plasma rotation. Analytical models predict that a modest amount of plasma rotation is sufficient to stabilise RWM growth \cite{Garofalo1999, Reimerdes2006, Sontag2005, Chu2010}; and both experiment and theory show that RWM drag can damp plasma rotation \cite{Garofalo1999,Garofalo2001a}.

Here, a model system of equations is derived containing linear RWM and nonlinear NTM models and extended to include self-consistent nonlinear interactions with toroidal plasma rotation. These equations are examined in a limiting case relevant to the tokamak and applied as a possible model for exploring the phenomenon of triggerless NTMs.
Section \ref{section:theoretical} will outline the derivation of the model system considered. This is followed by an analytic and numerical examination of the stability of the RWM-NTM coupled mode and its dependence upon model parameters in Section \ref{numerical}. Conclusions are discussed in Section \ref{conclusion}.

\section{Theoretical Model}\label{section:theoretical}
\begin{figure}[h!]
\begin{center}
\includegraphics[width=120mm]{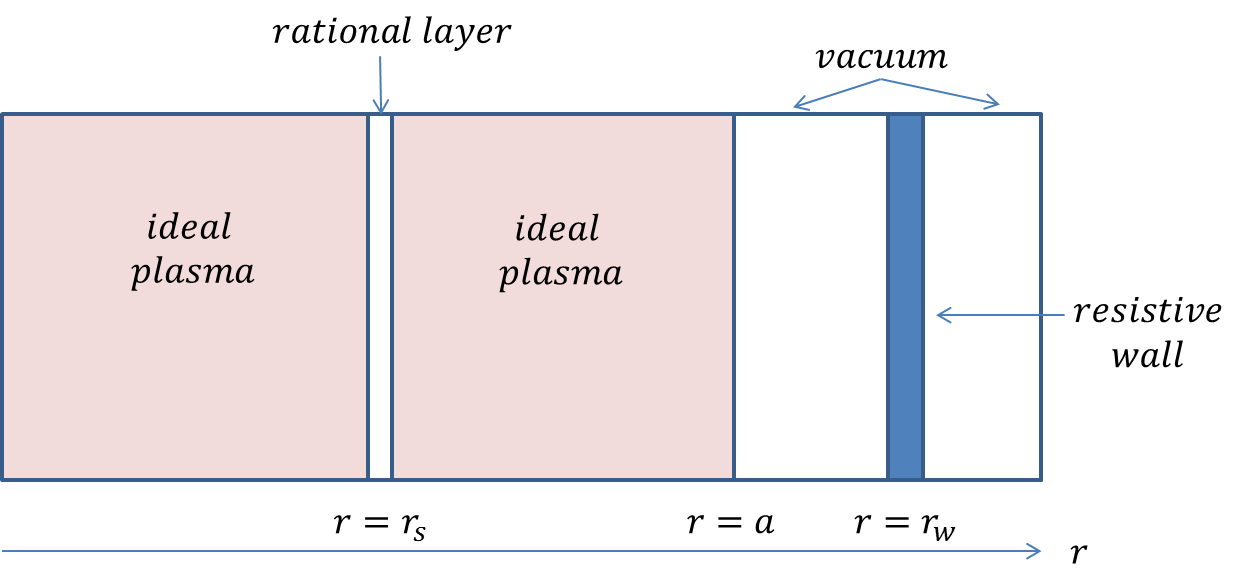}
\end{center}
\caption{The plasma is assumed to be ideal apart from the narrow layer near a rational surface (the ``rational layer'') where nonlinear effects and resistivity are important. The plasma is surrounded by a resistive wall and vacuum.}
\label{fig:schematic}
\end{figure}

Suppose an ideal, toroidally rotating plasma is surrounded by a thin wall at $r=r_w$ with finite conductivity $\sigma_w$, and contains a resistive layer in the vicinity of an internal rational surface at radius $r=r_s$ where a magnetic island can form. The regions of the plasma outside the resistive layer are assumed to be described by linear ideal MHD. Within the layer we retain resistivity and the nonlinear effects associated with magnetic islands. The plasma geometry is shown in Figure \ref{fig:schematic}. 
Due to skin currents in the resistive wall and at the rational surface, the component of the vector potential parallel to the equilibrium magnetic field has a discontinuous radial derivative at these locations. These discontinuities are parameterised by $\Delta_w$ and $\Delta_L$ respectively. In a linear, ideal MHD, cylindrical model employing a complex representation of the perturbed fields, they are related through an expression of the form \cite{Gimblett2004}
\begin{equation} \label{eq: dispersion}
\Delta_L =\frac{1- \delta \Delta_w }{-\epsilon + \Delta_w}   .
\end{equation} 
Here $\epsilon$ and $\delta$  depend on the equilibrium, and are related to the stability properties of the plasma in the limit of no wall and a superconducting wall respectively. They can be derived for a given equilibrium by solving ideal MHD equations outside the wall and rational layer with appropriate boundary conditions at $r=0$ and $r=\infty$, and noting the definition:
\begin{equation}
\Delta_{L,w}=\frac{r_{s,w}}{\psi_c} \frac{\partial \psi_c}{\partial r} \bigg |^{r_{s,w}^+}_{r_{s,w}^-}
\end{equation}
Here $\psi_c$ is the complex representation of the magnetic flux function derived from linear ideal MHD.
When no wall is present, $\Delta_w =0$ and for an ideally unstable plasma (where an inertial response for the layer is appropriate and $\Delta_L<0$ corresponds to instability to ideal MHD modes), $\epsilon$ is positive and small. For an ideal superconducting wall at $r=r_w$, $\Delta_w \rightarrow \infty$ and $\delta > 0$ provides stability to the internal kink mode, i.e. $\Delta_L<0$. Thus equilibria with small $\epsilon, \delta >0$ are susceptible to a RWM but avoids the internal kink mode. 
Whilst we solve linear equations in the outer region, we must anticipate a need to match to a nonlinear layer solution at the rational surface. Nevertheless, we assume we can separate the time dependence and write the outer region solution in the form:
\begin{equation}   \label{eq:magfluxcomplex}
\psi_c(r,t)=\tilde{\psi}(r)e^{i \xi} e^{\int p dt}
\end{equation}
where $p(t)=\gamma(t) - i \omega(t)$, with $\gamma$ the instantaneous growth rate, $\omega$ the mode frequency relative to the wall and $\xi=m\theta-n\phi$ the helical angle defined in terms of the poloidal ($\theta$) and toroidal ($\phi$) angles. It will be assumed that the toroidal mode number $n=1$, with $m$ the poloidal mode number.
Using Amp\`{e}re's Law, the wall response can be written as
\begin{equation} \label{eq: thin wall approximation}
 \Delta_w=p \tau
\end{equation}
where $\tau=\tau_w/\tau_r$ and $\tau_w=\mu_0 d \sigma_w r_w$ is the wall diffusion time, with the wall thickness $d<<r_w$, the minor radius at which the wall is located.
The {instantaneous eigenvalue} $p$ is normalised to $\tau_r=\sigma \mu_0 a r_s$ the resistive plasma diffusion time where $\sigma$ is the plasma conductivity at the rational surface.
Combining these results relates the layer response $\Delta_L$ to the wall properties and the instantaneous complex growth rate \cite{Gimblett2004}:
\begin{equation} \label{eq: linking}
\Delta_L =\frac{1- \delta \tau p }{-\epsilon +\tau p}
\end{equation}

In the linearised ideal MHD region, a complex representation for the magnetic flux is required to ensure that relation (\ref{eq: dispersion}) holds (a consequence of solving a second order ordinary differential equation for $\psi_c$ in this region). The physical flux is denoted $\psi_L=Re[\psi_c]$. In the rational layer where non-linear physics is retained, the complex representation cannot be used, so we work with this physical flux and assume a magnetic island exists at the rational surface. Let us transform into the frame of reference where the rational surface is at rest, so that the wall rotates.
Analogous to the wall, Amp\`{e}re's law describes the current perturbations
 \begin{equation}
 \frac{\partial^2 \psi_L}{\partial r^2}=-\mu_0 J_{||}
 \end{equation} 
where $J_{||}$ is the component of current density perturbation parallel to the magnetic field. There are two key contributions to $J_{||}$ to consider: an inductive component proportional to $\partial \psi_L/\partial t$ and the perturbation in the bootstrap current $J_{bs}$ caused by the pressure flattening inside the island. We assume that these current perturbations are localised within the layer and integrate across that layer to derive the model equation:
\begin{equation} \label{eq: surface int}
\frac{\partial \psi_L}{\partial r}\bigg|_{r_{s^-}}^{r_{s^+}} = 2\sigma \mu_0 w \frac{\partial \psi_L}{\partial t} - 2\mu_0 w {J_{bs}}
\end{equation} 
 where $w$ is the half-width of the magnetic island. The discontinuity on the left of Equation (\ref{eq: surface int}), caused by the current sheet in the layer, can be expressed in terms of $\Delta_L$ as follows. Let $\psi_L$ be of the form $\psi_L=\tilde{\varphi} cos(\hat{\alpha} + \zeta)$, where $\hat{\alpha}=\xi-\int \hat{\omega} dt$, {$\zeta(r)$} is a phase factor,  and the mode frequency in this rotating frame is $\hat{\omega}=\omega-\Omega_L$, where $\Omega_L$ is the toroidal plasma rotation frequency at $r=r_s$ relative to the stationary wall. 
Matching the ideal solution provides $Re[\tilde{\psi}] e^{\int \gamma dt}=\tilde{\varphi}cos(\Omega_L t+\zeta)$ and $Im[\tilde{\psi}] e^{\int \gamma dt}=\tilde{\varphi}sin(\Omega_L t+\zeta)$ (we assume that the toroidal mode number n=1 and $\tilde{\psi}$ is the amplitude of the magnetic flux function in Equation (\ref{eq:magfluxcomplex})). Thus, combining the results yields
\begin{eqnarray} \label{eq: }
&(\tilde{\varphi}/r_s) ( Re[\Delta_{L}] \cos(\hat{\alpha}+\zeta) - Im[\Delta_{L}] \sin(\hat{\alpha}+\zeta) )=&\nonumber \\
&2\sigma\mu_0 w( \frac{\partial \tilde{\varphi}}{\partial  t}  \cos(\hat{\alpha}+\zeta)+ \hat{\omega} \tilde{\varphi} \sin(\hat{\alpha}+\zeta) ) -2\mu_0 w J_{bs}&
\end{eqnarray}
Multiplying by $\cos(\hat{\alpha}+\zeta)$ or $\sin(\hat{\alpha}+\zeta)$ and then integrating over $\hat{\alpha}$ yields equations for ${\omega}$ and $\frac{\partial \tilde{\varphi}}{\partial  t}$. The island half width $w$ is related to $\tilde{\varphi}$ by
\begin{equation} \label{eq: island flux}
w^2=\frac{4 r_s L_s \tilde{\varphi}}{B_{\theta} q R}
\end{equation}
where {$L_s=\frac{R q}{s}$} is the shear length scale, $q$ is the safety factor, {$s$ is the magnetic shear,} $R$ is the major radius and $B_{\theta}$ is the poloidal magnetic field. Normalising the island half width $w$ to the minor radius, and time to $\tau_r$ we obtain 
\begin{eqnarray}
4 \dot{w} = Re[\Delta_{L}] +\frac{\hat{\beta}}{w}\bigg(1- \frac{w_c^2}{w^2}\bigg)  \label{eq:1}\\ 
 2 w(\omega -  \Omega_L)+Im[\Delta_{L}]=0	\label{eq:2}
\end{eqnarray}
where $\dot{w}$ refers to time derivatives with respect to $\hat{t}=t/\tau_r$ and $J_{bs}=\frac{\beta_{\theta} B_{\theta} {\sqrt{\hat{\varepsilon}}}}{L_p \mu_0} cos(\alpha+{\zeta})$, where $\beta_{\theta}$ is the poloidal beta, {$\hat{\varepsilon}$} the inverse aspect ratio,{ $L_p=\frac{1}{p} \frac{dp}{dr}$ the pressure length scale}, and $w_{c}$ the seed island threshold width. We have defined $\hat{\beta}=\frac{8 \sqrt{{\hat{\varepsilon}}} \beta_{\theta} r_s}{L_p s}$.
 Note that $\gamma=2 \dot{w}/w$, $\omega$ and $\Omega_L$ are normalised to $\tau_r$.  We have also introduced a heuristic NTM threshold factor $(1- w_c^2/w^2)$, which could  be attributed to the polarisation current effects in the plasma{ \cite{Wilson1996,Smolyakov1995}}. We have neglected geometrical factors to yield a simpler model which nevertheless retains the essential physics.  
Equation (\ref{eq:1}) is the same relation as that found in \cite{Wilson2000} when rotation is not considered.{ The full expression for $\Delta_L$ is given by
\begin{equation}\label{eq:delta}
\Delta_L=\frac{(1-\delta \tau \gamma)(-\epsilon+\tau\gamma)-\delta \tau^2 \omega^2}{(-\epsilon+\tau\gamma)^2+\tau^2\omega^2}+i\frac{\tau \omega (1-\epsilon \delta)}{(-\epsilon+\tau\gamma)^2+\tau^2\omega^2}
\end{equation}
}
To close the system, we require a torque balance relation. Ideal plasma is torque-free \cite{Taylor2003}; the torque exerted on the plasma is a delta function at the rational surface $\delta(r-r_s)$.

Consider the perturbed MHD momentum balance equation in the non-ideal layer, averaged over the flux surface.
\begin{equation} \label{eq:momentum conservation}
 \langle \mathbf{\delta} \mathbf{J} \times  \mathbf{\delta} \mathbf{B} \rangle + \rho\mu \nabla^2 \mathbf{v}=0
\end{equation}
where angled brackets denote the average.  Here $\rho$ is the plasma density and $\mu$ is the plasma viscosity. We assume that the processes under consideration occur over many viscous diffusion times, and pressure is constant across the thin layer. The pressure gradient is neglected as it will not contribute once we integrate across the layer (as follows).
Assume that $\mathbf{v}$ is continuous across the layer but $\nabla v$ possesses a discontinuity at $r=r_s$, due to the localised torque at that location.  For a perturbation $\delta \mathbf{B}$, $\nabla \times \delta \mathbf{B}=- \nabla^2 \psi_L \mathbf{b}$.
Only the toroidal component is required. Integrating Equation (\ref{eq:momentum conservation}) over the rational layer:
\begin{equation}
\bigg \langle \frac{\tilde{\varphi}}{R \mu_0}sin(\hat{\alpha}+\zeta) \frac{\delta \psi_L}{\delta r} \bigg |^{r_s^+}_{r_s^-} \bigg \rangle+ \rho \mu \frac{\delta v_{\phi}}{\delta r} \bigg | ^{r_s^+}_{r_s^-}=0
\end{equation}
The discontinuity in the radial derivative of $\psi_L=Re[\psi_c]$ is evaluated as above. Integrating over $\hat{\alpha}$, the terms in $Re[\Delta_{L}]$ disappear. Let  $v_{\phi}=\Omega R$ and normalise as before to derive
\begin{equation} \label{eq:torque balance}
\frac{\delta \Omega}{\delta r} \bigg|^{r_{s}^{+}}_{r_{s}^{-}}= \frac{A}{a} w^4 Im[\Delta_L]
\end{equation}
such that $A=s^2 {\hat{\varepsilon}}^2 a^3 \tau_V \tau_r/512  r_{s}^3 q^2 \tau_A^2$,with $\tau_V=a^2/\mu$ the momentum confinement time, $a$ the minor radius,  and {$\tau_A=\frac{a \sqrt{\mu_0 \rho}}{B}$ the Alfv\'{e}n time} and taking the $\sigma=T_e^{3/2}/1.65 \times 10^{-9} \ln \lambda$, with $T_e$ expressed in keV. {The safety factor $q$ at $r=r_s$ will be taken as $q=2$.}
The electromagnetic torque is finite when $Im[\Delta_L] \neq 0$, and always acts to damp the plasma rotation \cite{Fitzpatrick2002}. $Im[\Delta_L]$ is related to the discontinuity in the derivative of the phase factor $\zeta$ \cite{Gimblett2004}: the torque only acts on the plasma in the rational layer when $\frac{\partial \zeta}{\partial r} |_{r_s^-}^{r_s^+} \neq 0$.

 In the outer ideal plasma regions, the momentum equation is simply
\begin{equation}
\frac{d^2 \Omega}{d r^2}=0
\end{equation}
The linear rotation profile in the ideal plasma regions, $r<r_s$ and $r_s<r<a$, is continuous at $r=r_s$ but has a discontinuity in the first radial derivative there, caused by the torque. Imposing no slip boundary conditions at the outer edge of the plasma $r=a$ (viscous drag effects) and a driving force at the inner edge at $r=0$ that maintains $\frac{d\Omega}{dr} |_{r=0}=-\Omega_0/a$, the toroidal rotation frequency profile can be constructed in the region of ideal plasma. 
\begin{eqnarray}
\Omega (r)&=&\lambda (a-r_s)+\frac{\Omega_0}{a}(r_s -r)\quad        0< r < r_s\\
\Omega (r)&=&\lambda (a-r)      \quad  \quad\quad\quad  \qquad \quad  r_s<r<a
\end{eqnarray}
for constant $\lambda>0$.
The discontinuity in the radial derivative is calculated to be $-\lambda + \Omega_0/a$.
Hence, at $r=r_s$, using Equation (\ref{eq:torque balance}), the damping of the toroidal plasma rotation frequency is found to be
{
\begin{equation} \label{eq:rational omega value}
(\Omega_0-f \Omega_L)=A w^4 Im[\Delta_L]
\end{equation}
}
where $f=\frac{a}{a-r_s}$.

{
\section{Solutions For Coupled RWM-NTM Mode Stability}\label{numerical}

The full system of nonlinear equations (\ref{eq: linking}), (\ref{eq:1}), (\ref{eq:2}), and (\ref{eq:rational omega value}) can be solved numerically and evolved in time, with {$\tau\approx 10^{-2}$ corresponding to the choice of $a=1.0m, r_s=0.5m,T_e=1keV, r_w=1.0m$, and Coulomb logarithm  $ln\Lambda=20$. We take $\epsilon=0.1, \delta=5$}.

If there is no wall surrounding the plasma, then $\tau=0$. In this situation, the RWM is not a solution (an infinite growth rate is obtained, which can be interpreted as the ideal kink mode). However, the NTM is a solution in this limit (Figure \ref{fig:tau0}), exhibiting its characteristic features: specifically a sufficiently high $\hat{\beta}$ must be achieved for the NTM to become unstable, as well as a seed island with a width $w>w_c$. An NTM will saturate at a large island width, proving detrimental to plasma confinement.

 \begin{figure}
\begin{center}
\includegraphics[width=160mm]{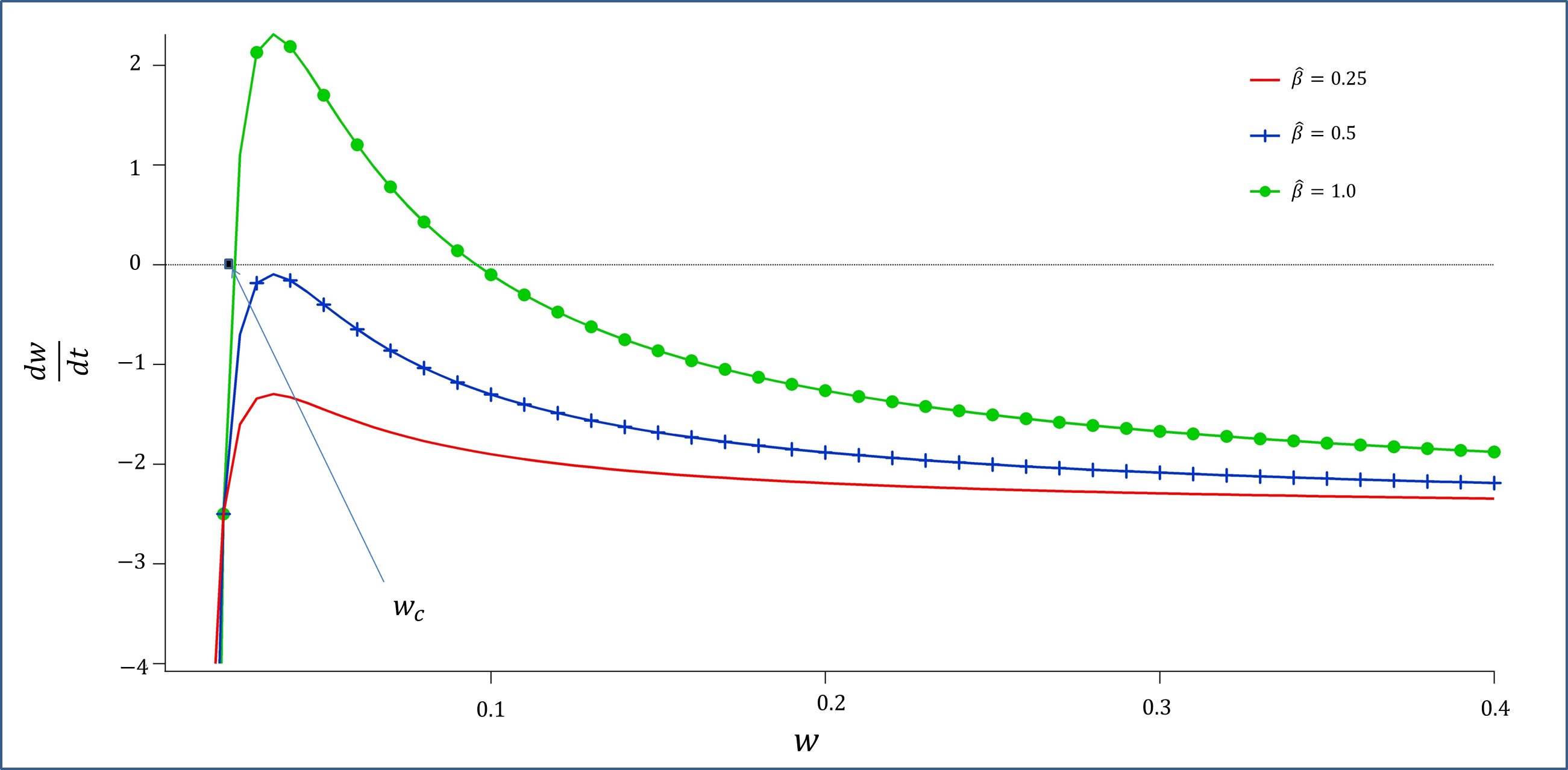}
\end{center}
\caption{No RWM exists when the resistive wall is removed: only the NTM is a solution. The dependence of the island width evolution on $\hat{\beta}$ and $w_c$ illustrates the behaviour expected for a NTM.}
\label{fig:tau0}
\end{figure} 

When $\tau$ is finite, both RWM and NTM branches are present. If the wall is thin, or has a low conductivity then $\tau <<1$ and we can solve the equations analytically, as follows.

\subsection{Analytical Solutions For Small Islands}\label{analytic}

We first consider the behaviour of the solutions when $w<w_c$. For a standard NTM, an island of this size would not grow as it is smaller than the threshold width. The relevant ordering is $\tau \omega<<\tau \gamma \sim \epsilon$. The bootstrap term ($\sim \hat{\beta}$) in Equation (\ref{eq:1}) is relatively large at small $w<w_c$ and negative, and this must be balanced by either a large $\dot{w}$ (i.e. $\gamma$) on the left hand side of Equation (\ref{eq:1}) or by a large value of $Re[\Delta_L]$. Neglecting $\tau \omega$ in Equation (\ref{eq:1}), the branch when $\gamma$ is large (and negative) is the standard stable NTM root of the equations-as illustrated in Figure \ref{fig:tau0}. Specifically, in this limit $Re[\Delta_L]\sim -\delta$ and $\gamma<0$ for $w<w_c$. The alternative is the RWM solution branch, in which the denominator of $Re[\Delta_L]$ is small. Neglecting the left hand side of Equation (\ref{eq:1}), and balancing $Re[\Delta_L]$ against the bootstrap term yields a RWM which grows at the rate:

\begin{equation}\label{eq:gamma_estimate}
\tau \gamma = \epsilon  -\frac{w}{\hat{\beta}}\Big(1-\frac{w_c^2}{w^2}\Big)^{-1}
\end{equation}

Thus, an island such that $w<w_c$ will have $\gamma >0$ and will grow steadily despite being below the NTM threshold width. When $w$ approaches $w_c$, the growth rate is substantially enhanced by the coupling to the bootstrap term. Indeed, as $w \rightarrow w_c$, $\tau \gamma \rightarrow \infty$, but this is unphysical and the ordering then breaks down. Similarly, for the assumptions that $\omega<<\Omega_L$ and $\tau \omega<<1$, we find the RWM frequency and toroidal rotation frequency:
\begin{eqnarray}
\label{eq:omega_estimate}
\tau \omega&=&\frac{2 w^3 \Omega_0}{(1-\epsilon \delta) \hat{\beta}^2}\Big( \frac{w_c^2}{w^2}-1\Big)^{-2}(f+2 A w^5)^{-1}\\ 
\Omega_L&=&\frac{\Omega_0}{f+2 A w^5} \label{eq:rot_estimate}
\end{eqnarray}

We see from Equation (\ref{eq:rot_estimate}) that the plasma rotation at the rational surface $\Omega_L$ decreases steadily in time from its initial value $\Omega_0/f$ as the island grows.  The mode is initially locked to the wall ($\tau \omega = 0$ when $w=0$), but as the island grows, the mode frequency $\omega$ gradually increases. Nevertheless, it remains very small, $<<\Omega_L$. This is because Equation (\ref{eq:gamma_estimate}) for the growth rate means $Im[\Delta_L]$, and therefore the electromagnetic torque, is very large. A consequence of this is that there is no rotational stabilisation of the RWM exhibited in Equation (\ref{eq:gamma_estimate}). This is different to the model presented in \cite{Gimblett2000} where the bootstrap term is not included. Then the RWM mode frequency rises faster with $\Omega_L$, leading to a suppression of its growth rate as $\Omega_L$ increases.

Note that $\omega$ is increasingly sensitive to $w$ as $w$ approaches $w_c$, indicating unlocking of the island and a dramatic spin-up. Indeed, as $w \rightarrow w_c$, Equation (\ref{eq:omega_estimate}) predicts $\tau \omega \rightarrow \infty$ and the ordering is again broken.
To recap, for $w<w_c$ we find a slowly growing mode which is locked to the wall- the RWM. As the island width $w$ approaches $w_c$, there is a substantial increase in the growth of the island, and the mode begins to spin up. The small $\tau \gamma$, small $\tau \omega$ ordering then fails.
Note from Equation (\ref{eq:delta}) that $Im[\Delta_L]\rightarrow 0$ as $\tau \gamma$ and $\tau \omega$ grow. Thus, the electromagnetic torque which locks the mode to the wall is reduced and the mode will spin up to rotate with the plasma. We see from Equation (\ref{eq:rational omega value}) that this forces $\Omega_0 - f \Omega_L \approx 0$, so the plasma also spins up towards the initial rotation frequency profile. In Figure \ref{fig:estimates}, we compare our analytic solutions in Equations (\ref{eq:gamma_estimate}), (\ref{eq:omega_estimate}) and (\ref{eq:rot_estimate}) to numerical solutions of the full system, Equations (\ref{eq:1}), (\ref{eq:2}) and (\ref{eq:rational omega value}). There is good agreement until $w$ approaches $w_c$, and then (as expected) the analytic theory fails. Nevertheless, there is qualitative agreement, both approaches showing a stronger growth and mode spin-up of the coupled RWM-NTM system as $w$ passes through $w_c$. In the following subsection, we employ our numerical solution to explore the coupled RWM-NTM mode in more detail.

 \begin{figure}[h!]
\begin{center}
\includegraphics[width=110mm]{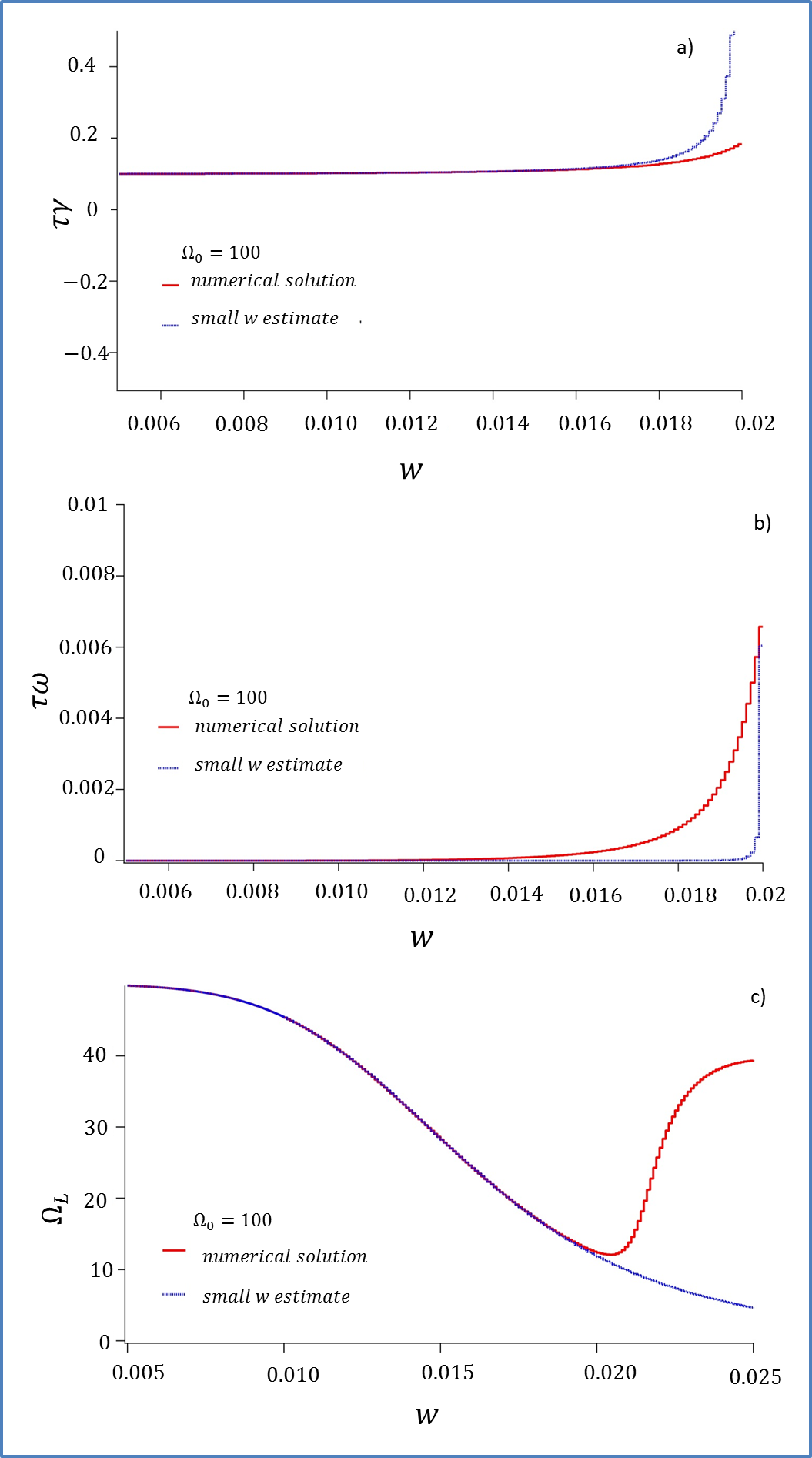}
\end{center}
\caption{The comparison of the analytic solutions valid for $w<w_c$ with the numerical solutions of the full system: a) and b) show both $\tau \gamma$ and $\tau \omega$ increasing rapidly as $w$ passes through $w_c$. In c), the forced plasma spin up observed in the numerical results is not captured in our analytic approximation. We have fixed $\Omega_0=100, w_c=0.02$ and $\hat{\beta}=1.0$}
\label{fig:estimates}
\end{figure}

\subsection{Dependence Of Numerical Solutions On $\hat{\beta}$ and $\Omega_0$}

In the case with zero $\hat{\beta}$, the NTM has no drive and we expect a conventional RWM: a mode that rotates at a fraction of the plasma rotation, but acting to slow the plasma, eventually locking to the wall. Rotation is expected to be stabilising in this situation \cite{Gimblett2000}. We have seen in the previous subsection both analytically and numerically that at sufficiently large $\hat{\beta}$ the RWM couples to an NTM, and takes on a different character. In this subsection we now explore the essential physics of this coupling, considering the situation where a seed island has a width smaller than the NTM threshold width $w_c$, such that the NTM solution is stable (i.e. $w_{seed}=w_c/4$). 

Numerical solutions are shown for a range of $\hat{\beta}$ in Figure \ref{fig:beta}. We observe no threshold seed island width for this coupled mode: a seed island of any size is able to grow. In Figure \ref{fig:beta}a), the seed island grows steadily independent of $\hat{\beta}$ as expected for a ``classic'' RWM until it reaches width $w_c$. This is consistent with the analytic solution, Equation (\ref{eq:gamma_estimate}), $\gamma=\epsilon/\tau$. As $w$ increases above $w_c$, the island growth rate increases dramatically for high $\hat{\beta}$ as the bootstrap term then becomes destabilising, providing an additional drive for the mode (this cannot be captured in the analytic results). If $\hat{\beta}$ is reduced, then the NTM drive is reduced and its effect on the island growth is either reduced or insignificant. The plasma rotation frequency at $r=r_s$, $\Omega_L$, is damped by torque exerted by the unstable RWM, until the island width reaches $w_c$ (Figure \ref{fig:beta}c)). The plasma then briefly spins up as the RWM couples to the NTM drive, before again slowing to again lock to the wall at large $w$. Note that throughout the period for which $w \geq w_c$, the plasma and mode are locked- the mode rotates with the plasma. The magnitude of the spin up depends on $\hat{\beta}$- increasing the NTM drive clearly affects the RWM branch. The transient spin up takes place over a time scale of 3-4 ms for the parameters we have chosen. (This may invalidate the assumption that inertia can be neglected.) Decreasing $\hat{\beta}$ causes the rotation spin up to reduce and eventually be removed, consistent with the picture that the NTM drive is the underlying physics. It can be seen in Figure \ref{fig:beta} that removing the NTM drive by reducing $\hat{\beta}$ causes the solution to behave more as a classic RWM, with wall locking of both the mode and plasma.

 \begin{figure}[h!]
\begin{center}
\includegraphics[width=200mm, angle=90]{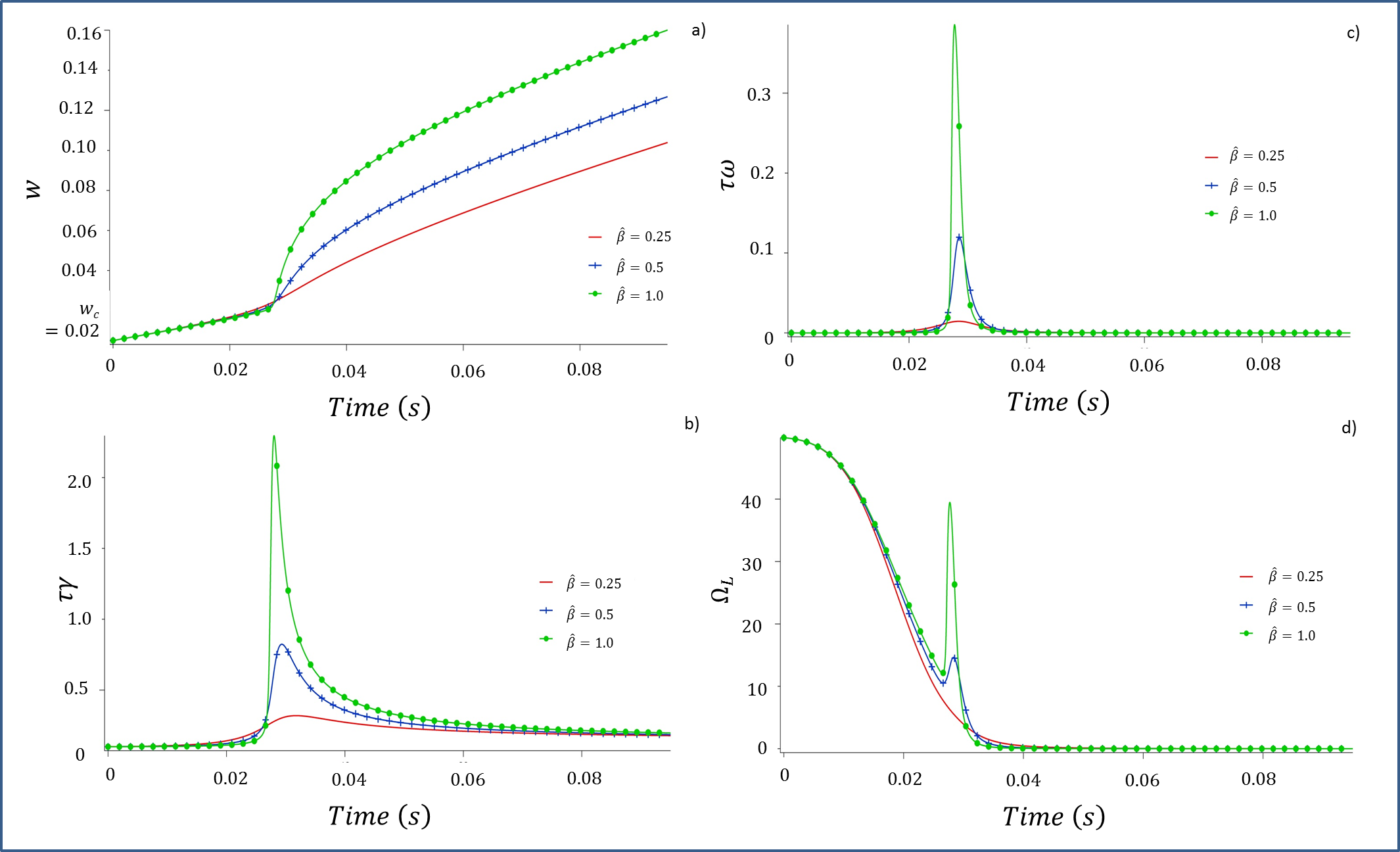}
\end{center}
\caption{Here $w(t=0)=0.005$ which is considerably smaller than the critical seed island width $w_c=0.02$. The dependence of the island growth, growth rate $\gamma$, mode frequency and plasma rotation frequency at the rational surface on $\beta$ is shown: increasing $\beta$ (which increases the NTM drive) allows the island to grow faster and to a larger size.}
\label{fig:beta}
\end{figure} 

The island evolution is also influenced by the amount of momentum injected into the plasma core, as characterised by $\Omega_0$. This is shown in Figure \ref{fig:omega0}. Note that for our model the RWM is not sensitive to the plasma rotation, as evidenced by the independence from $\Omega_0$ of the initial evolution of $w$ until $w=w_c$. This is consistent with our analytic solution in Equation (\ref{eq:gamma_estimate}) but differs for models \cite{Gimblett2000} that do not include the NTM drive. The difference appears to be because the coupled system has a stronger electromagnetic torque and the mode is locked to the wall for even large rotation frequencies. It is a rotation of the mode relative to the wall that provides the stabilisation in \cite{Gimblett2000}. We find that the coupling to the NTM is strongly influenced by flow, and the dramatic increase of island growth rates for $w>w_c$ is only observed at the lowest $\Omega_0$. At higher $\Omega_0$, the mode evolves as a classic RWM which is locked to the wall, with a substantially reduced growth rate compared to the low $\Omega_0$ cases when $w>w_c$.

\begin{figure}[h!]
\begin{center}
\includegraphics[width=165mm, angle=90]{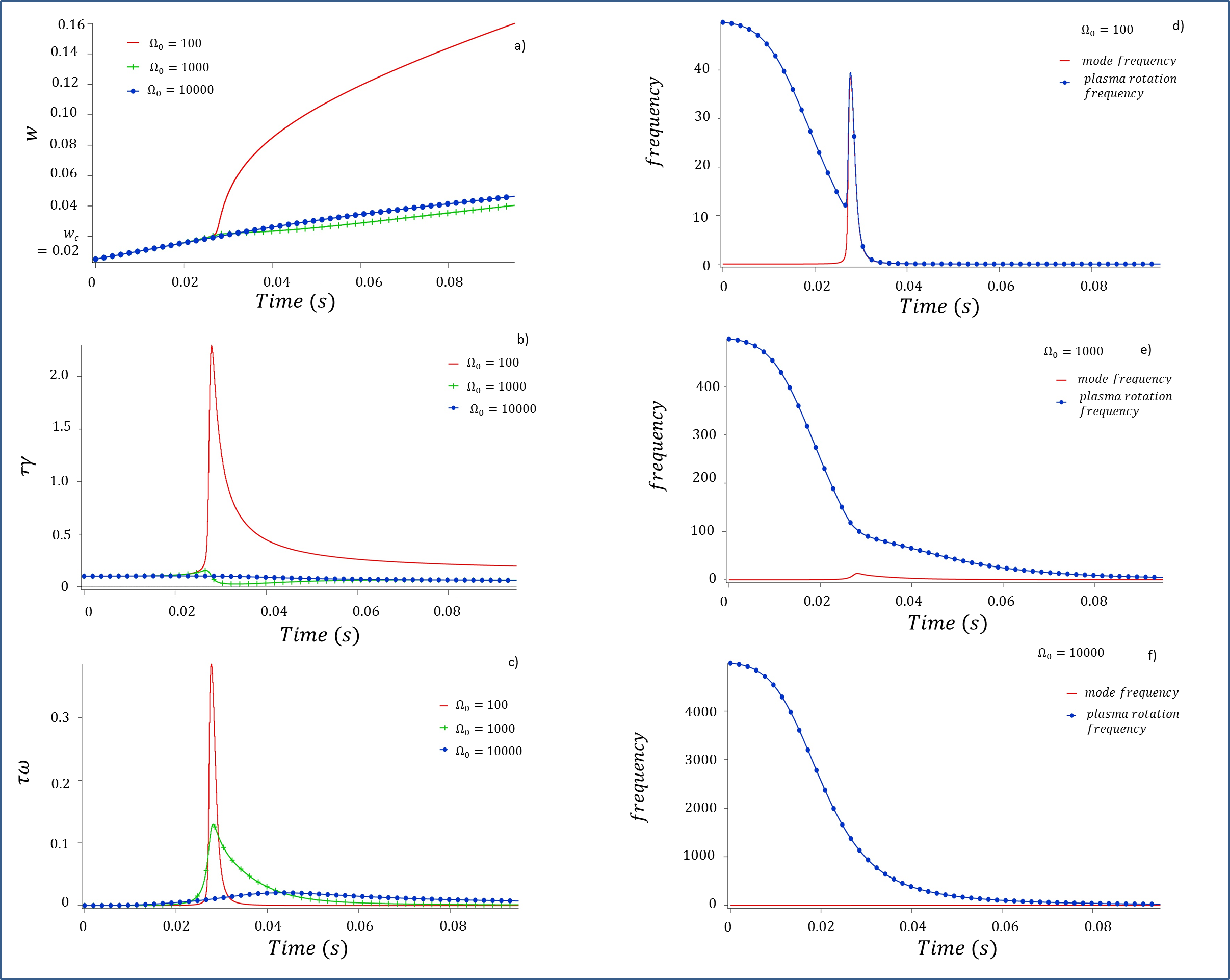}
\end{center}
\caption{$w(t=0)=0.005, \hat{\beta}=1.0$ a) demonstrates that the initial plasma rotation frequency has a noticeable effect on the island evolution, whereas b) and c) contrast the mode growth rates and mode frequencies for differing values of $\Omega_0$.  d), e) and f) show the mode and rotation frequencies for the same values of $\Omega_0$. $\Omega_0=10000$ corresponds to the plasma rotating at about $5\%$ of the plasma sound speed- this spin up only occurs for very small rotation frequencies.}
\label{fig:omega0}
\end{figure}

\section{Conclusion}\label{conclusion}

In this paper, we have developed and analysed a simple model for the coupling between a RWM and an NTM in tokamaks, retaining the effects of plasma rotation. We find two branches- a mode which is essentially an NTM with a threshold, and a coupled RWM-NTM which has no threshold.

While the RWM-NTM island width is small (below the NTM threshold), the mode is a classic RWM, growing slowly (on a timescale characterised by the wall resistive diffusion time) and locked to the wall. The plasma rotation gradually slows during this phase. As this mode has small amplitude ($w \sim 1$cm) and is locked, it would be difficult to detect in an experimental situation. As the island width exceeds the NTM threshold, there is a dramatic increase in growth, particularly at high $\beta$ as the mode couples to the NTM drive. At the same time, the mode unlocks from the wall and instead rotates with the plasma. The plasma rotation also increases at this time. The island continues to grow, locked to the plasma flow, which gradually slows to lock to the wall at large island width. This phase has the characteristics of an NTM.

Although the early evolution of the coupled RWM-NTM would be difficult to detect experimentally, once the island width exceeds the NTM threshold $w_c$, and the mode spins up, it would then be detectable by Mirnov coils. Indeed, at this time it would already have a width $w\sim w_c$ ($\sim 1$cm) and so would have the appearance of a triggerless NTM- a $\sim 1$cm island which does not appear to come from a seed island.

The phenomenon of triggerless NTMs near $\beta_N^{no-wall}$, the ideal no-wall $\beta$ limit,  can also be explained by invoking a pole in $\Delta^{'}$ (which here is equal to $\Delta_L/r_s$). As $\beta_N \rightarrow \beta_N^{no-wall}$, $\Delta^{'} \rightarrow \infty$, destabilising a classical tearing mode, which in turn destabilises the neoclassical tearing mode when at a sufficiently large amplitude \cite{Brennan2003}. In our model, this would be equivalent to letting $\Delta_w \rightarrow 0$ and $\epsilon \rightarrow 0$, corresponding to the no-wall ideal MHD stability boundary. In the model described here, the RWM is destabilised for $\beta_N>\beta_N^{no-wall}$ but the pole in $\Delta_L$ is resolved by a dependence on the wall response. The physics interpretation is therefore different to that provided in \cite{Brennan2003}. Our model has some properties that can be tested experimentally, distinguishing it from the model with a pole in $\Delta^`$. First, the plasma is expected to slow in the few $10$'s of milliseconds before the island width reaches $w_c$. Then one would observe a dramatic spin-up of the plasma, coincident with the appearance of the mode on Mirnov coils, which would only last a few milliseconds before the plasma and the mode slow and lock to the wall.

The emphasis of this paper is directed towards understanding the importance of the RWM-NTM coupling in the presence of plasma rotation, not the physics details of each mode. Thus, while we have retained the essential physics, we have neglected more subtle effects such as the physics of rotation on the matching of the layer solution to the external, ideal MHD solution \cite{Shiraishi2011}. Future work should extend our model to include a more complete description of the RWM and NTM physics in order to make quantitative predictions for the behaviour of this couple RWM-NTM.

\section{Acknowledgments}
The support of the Engineering and Physical Sciences Research Council for the Fusion Doctoral Training Network (Grant EP/K504178/1) is gratefully acknowledged. This work was funded partly by the RCUK Energy Programme under grant EP/I501045 and the European Communities under the contract of Association between EURATOM and CCFE, and partly by EPSRC Grant EP/H049460/1. The views and opinions expressed herein do not necessarily reflect those of the European Commission.


\begin{thebibliography}{99}
{
\bibitem{ITER1999}
  ITER Physics Basics Editors \emph{et al.},
  Nucl. Fusion
  \textbf{39},
  2137
  (1999).
  }
\bibitem{Kessel1994}
C. Kessel, J. Manickham, G. Rewoldt, and W. M. Tang,
  Phys. Rev. Lett.
  \textbf{72},
  1212
  (1994).
    \bibitem{Carrera1986}
  R. Carrera, R. D. Hazeltine and M. Kotschenreuther,
  Phys. Fluids
  \textbf{29},
  899
  (1986).
{
\bibitem{Hender2007}
T. C. Hender \emph{et al.},
\emph{Progress in the ITER Physics Basis Chapter 3: MHD stability, operational limits and disruptions (beginning of chapter)}
  Nucl. Fusion
  \textbf{47},
  S128-S202
  (2007).
  }
\bibitem{Gerhardt2009}
S. Gerhardt \emph{et al.},
Nucl. Fusion
\textbf{49},
032003
(2009).
\bibitem{Haye2006}
R. J. La Haye,
  Phys. Plasmas
  \textbf{13},
  055501
  (2006).
  \bibitem{Sauter2002}
O. Sauter \emph{et al.},
Phys. Rev. Lett.
\textbf{88},
105001
(2002).
\bibitem{Chapman2010}
I. T. Chapman \emph{et al.},
Nucl. Fusion
\textbf{50},
102001
(2010).
  \bibitem{Haye2000a}
  R. J. La Haye,  R. J. Buttery, S. Guenter,  G. T. A. Huysmans,  M. Maraschek, and H. R. Wilson,
  Phys. Plasmas
  \textbf{7},
  3349
  (2000).
 \bibitem{Gude1999a}
 A. Gude, S. Guenter, S. Sesnic and {ASDEX}-{U}pgrade Team,
  Nucl. Fusion
  \textbf{39},
  127
  (1999). 
 \bibitem{Haye2000b}
 R. J. La Haye, B. W. Rice, and E. J. Strait,
  Nucl. Fusion
  \textbf{40},
  53
  (2000). 
   \bibitem{Chang1995}
   Z. Chang, \emph{et al.},
    Phys. Rev, Lett.
  \textbf{74},
 4663
  (1995).              
                  \bibitem{Fitzpatrick1995}
   R. Fitzpatrick,
  Phys. Plasmas
  \textbf{2},
  3 
  825
  (1995). 
    \bibitem{Wilson1996}
  H. R. Wilson, J. W. Connor, R. J. Hastie, and C. C. Hegna,
 Phys. Plasmas 
  \textbf{3},
  248
  (1996).   
  \bibitem{Kislov2001}
  D. A. Kislov \emph{et al.},
  Nucl. Fusion
  \textbf{41},
  1619
  (2001). 
     \bibitem{Brennan2003}
  D. P. Brennan \emph{et al.},
  Phys. Plasmas
  \textbf{10},
  1643
  (2003). 
  \bibitem{Garofalo1999}
  A. M. Garofalo \emph{et al.},
  Phys. Rev, Lett.
  \textbf{82},
  3811
  (1999). 
  {
      \bibitem{Maget2010}
  P. Maget \emph{et al.},
  Nucl. Fusion
  \textbf{50},
  045004
  (2010).
  } 
 \bibitem{Reimerdes2006}
  H. Reimerdes \emph{et al.},
  Phys. Plasmas.
  \textbf{13},
  056107
  (2006). 
   \bibitem{Sontag2005}
  A. C. Sontag \emph{et al.},
  Phys. Plasmas.
  \textbf{12},
  056112
  (2005). 
  \bibitem{Chu2010}
  M. S. Chu and M. Okabayashi,
  Plasma Phys. Contr. Fusion
  \textbf{52},
  123001
  (2010). 
   \bibitem{Garofalo2001a}
   A. M. Garofalo and et al.,
  Nucl. Fusion
  \textbf{41},
  1171
  (2001).   
  \bibitem{Gimblett2004}
  C. G. Gimblett and R. J. Hastie,
  Phys. Plasmas
  \textbf{11},
  3,
  1019
  (2004). 

      \bibitem{Smolyakov1995}
A. I. Smolyakov and et al.,
 Phys. Plasmas 
  \textbf{2},
  1581
  (1995). 
    \bibitem{Wilson2000}
    H. R. Wilson, J. W. Connor, C. G. Gimblett , R. J. Hastie, and F. L. Waelbroeck,
  Proc. 18th Int. Conf. Sorrento, 2000, IAEA, Vienna,
  (2000). 
  \bibitem{Taylor2003}
  J. B. Taylor,
  Phys. Rev. Lett.
  \textbf{91},
  11,
  115002
  (2003). 
  \bibitem{Fitzpatrick2002}
  R. Fitzpatrick,
  Phys. Plasmas
  \textbf{9},
  3459
  (2002). 
    \bibitem{Gimblett2000}
   C. G. Gimblett and R. J. Hastie,
  Phys. Plasmas
  \textbf{1},
  258
  (2000). 
  
     \bibitem{Shiraishi2011}
  J. Shiraishi and S. Tokuda,
  Nucl. Fusion
  \textbf{51},
  053006
  (2011). 
\end{thebibliography}
\end{document}